# Velocity- and pointing-error measurements of a 300 000-r/min self-bearing permanent-magnet motor for optical applications

Sven Breitkopf[1,*], Nikolai Lilienfein[2,3], Timon Achtnich[4], Christof Zwyssig[4], Andreas Tünnermann[1,5,6], Ioachim Pupeza[2], and Jens Limpert[1,5,6]

1 - Institute of Applied Physics, Abbe Center of Photonics, Friedrich-Schiller-Universität Jena, Albert-Einstein-Str. 15, 07745 Jena, Germany
2 - Max-Planck-Institute of Quantum Optics, Hans-Kopfermann-Str. 1, 85748 Garching, Germany
3 - Ludwig-Maximilians-Universität München, Am Coulombwall 1, 85748 Garching, Germany
4 - Celeroton Ltd., Industriestrasse 22, 8604 Volketswil, Switzerland
5 - Fraunhofer Institute for Applied Optics and Precision Engineering, Albert-Einstein-Str. 7, 07745 Jena, Germany
6 - Helmholtz-Institute Jena, Fröbelstieg 3, 07743 Jena, Germany
* corrseponding author: sven.breitkopf@uni-jena.de

**Abstract**

**Compact, ultra-high-speed self-bearing permanent-magnet motors enable a wide scope of applications including an increasing number of optical ones. For implementation in an optical setup the rotors have to satisfy high demands regarding their velocity and pointing errors. Only a restricted number of measurements of these parameters exist and only at relatively low velocities. This manuscript presents the measurement of the velocity and pointing errors at rotation frequencies up to 5 kHz. The acquired data allows to identify the rotor drive as the main source of velocity variations with fast fluctuations of up to 3.4 ns (RMS) and slow drifts of 23 ns (RMS) over ~120 revolutions at 5 kHz in vacuum. At the same rotation frequency the pointing fluctuated by 12 µrad (RMS) and 33 µrad (peak-to-peak) over ~10000 roundtrips. To our best knowledge this states the first measurement of velocity and pointing errors at multi-kHz rotation frequencies and will allow potential adopters to evaluate the feasibility of such rotor drives for their application.**

*Index Terms*: **rotors, magnetic bearing, self-bearing, high-speed drives, slotless machine, ultrafast optics, lasers**

## I. Introduction

Over the last years, compact, ultra-high-speed self-bearing permanent-magnet motors have enabled a multitude of applications, such as turbo compressor systems, turbo molecular pumps, ultracentrifuges, and milling [1],[2]. They can currently run at rotation frequencies up to 500 000 revolutions per minute (rpm) [3],[4] with room for further improvement [5]. These unprecedented rotation frequencies as well as the running smoothness and longevity afforded by such motors are of great promise for application in optical devices such as rotating polygon-scanners used in material processing, 3D measurements, and wavelength-swept lasers [6],[7] for biomedical applications. In the context of high pulse-energy, high repetition-rate ultrafast lasers, the use of such motors in optical switches for stack-and-dump cavities [8],[9] and regenerative amplifiers [10],[11] has recently been proposed [12]. Such optical applications are often highly sensitive to the pointing stability and velocity errors of the rotor, which translate directly to the precision and reproducibility of the beam deflection. Precise measurements of these properties at ultra-high velocities have not been published to date. Publications of measurements at slower rotation frequencies suggest that the relative velocity error can be well below 0.02 % at 4 000 rpm (~67 Hz) and above [13]. In this manuscript measurements of the velocity error of a customized self-bearing permanent-magnet motor are presented for a range of rotation frequencies up to 300 000 rpm (5 kHz) in air and in vacuum. The results show fast fluctuations of the rotation period of 3.4 ns (RMS), 0.0017% respectively and a slow drift of at least 23 ns (RMS), 0.013 % respectively at a rotation frequency of 5 kHz over ~120 revolutions in vacuum, and fast fluctuations of the rotation period of 5.2 ns (RMS) and a slow drift of at least 33.3 ns (RMS) at a rotation frequency of 4 kHz over ~40 revolutions in air. The velocity error in air is found to be limited by air fluctuations, while the performance in vacuum could potentially be improved by optimizing the motor drive unit. Additionally, the asynchronous pointing stability of the rotor was investigated, measuring fluctuations of 12 µrad (RMS) and 33 µrad (peak-to-peak) over ~10000 roundtrips at 5 kHz in vacuum. This states the first systematic investigation of a self-bearing permanent-magnet motor operating at multi-kHz frequencies in terms of rotation-period jitter and asynchronous pointing jitter.

## II. Design and properties of the motor

The device under test consists of two active magnetic radial bearings, an active magnetic axial bearing and the motor drive is integrated into a heteropolar and a homopolar motor, allowing for a rotor with a load in the axial center of the device. The rotor sleeve is made of titanium and contains one radially magnetized permanent magnet (PM) for the first radial bearing and the motor drive and two axially magnetized PMs for the second radial bearing and the axial bearing [14]. For the purpose of the measurements presented herein, the motor is equipped with a customized rotor, containing a triangularly cross-sectioned volume located in the axial center of the rotor. One facet of the volume has a polished surface in order to reflect a laser beam to enable the desired measurements (see Fig. 1).

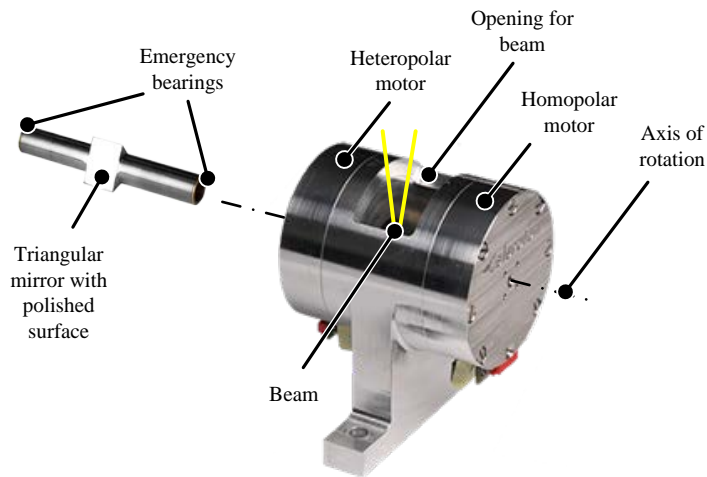

Fig. 1 Employed rotor (left) and entire magnetic bearing motor (right). One of the three rotor mirror-facets is polished in order to reflect a laser beam.

## III. VELOCITY ERROR MEASUREMENTS

The rotation period jitter $\Delta t$ of the revolution $n$ is defined as the deviation of the rotation period $t_n$ from the average rotation period $t_{avg}$. To allow for meaningful statements, it is necessary to measure the jitter of every revolution over a certain number of subsequent revolutions with a relative accuracy of $\Delta t/t_{avg} \approx 1$ ppm (part-per-million). In order to measure this very small deviation with a simple optical setup consisting of a continuous-wave laser (1064 nm central wavelength), the metal rotor with a polished surface, and a fast photodiode (see Fig. 2).

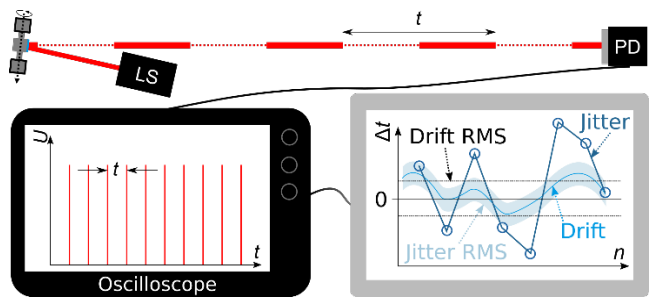

Fig. 2 LS – *Laser*: Mephisto (S) Non-planar Ring Oscillator (continuous wave, λ = 1064 nm). PD – *Photodiode*: PDA10CF-EC, *Oscilloscope*: LeCroy WaveMaster 830Zi-B.

The laser beam reflected from the polished facet of the rotor hits the photodiode (PD) at a distance $s = 2$ m from the rotor once during every rotation period. For the measurements in vacuum the motor was placed in a vacuum chamber with a residual pressure of around 0.7 mbar, while the light source and detectors remained in an air environment. For these measurements the distance was increased to $s = 3$ m. The oscilloscope acquires a trace containing the signal from a certain number of revolutions (Fig. 3a). For the measurements in air, the oscilloscope saved traces of up to $16 \times 10^6$ samples at rates of up to $30 \times 10^9$ samples/s. This was further optimized for the later carried out measurements in vacuum, allowing to save up to $64 \times 10^6$ samples. To record as many revolutions as possible without compromising the measurement accuracy, the varying sample rates are chosen to achieve between $0.2 \times 10^6$ and $3 \times 10^6$ samples per revolution. The resulting time traces contain between 16 and 50 revolutions for the measurements in air, and between 20 and 320 revolutions for the vacuum measurements. The longer traces give better information about slow fluctuations, while the shorter traces offer a higher temporal resolution, which is particularly useful for the fast rotation speeds. Before extraction of the rotation periods from the acquired oscilloscope traces, two processing steps are carried out.

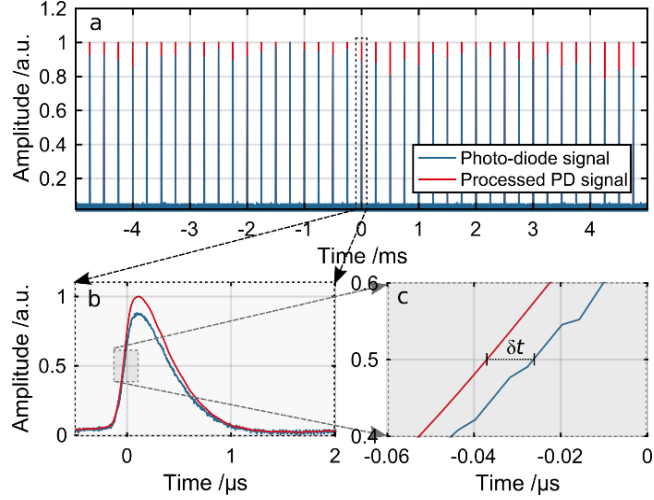

Fig. 3 Post-processing of the acquired traces permitting to accurately calculate the time delay between the roundtrips. The main goal is to remove the temporal error δt otherwise occurring when the trace is directly used without any post-processing.

First, the signal is filtered employing a low-pass Fourier-filter with a hard-cut at 10 MHz to dispose of the noise on the PD signal, which could potentially cause problems for the rising-edge detection (see Fig. 3b and c). Second, since the peaks vary in terms of amplitude due to beam pointing fluctuations between the revolutions (see next section) each peak is normalized to its maximum (visible in Fig. 3a). This step is important in order to avoid timing errors δt when scanning for the temporal position of each revolution (see Fig. 3c). An algorithm detecting the rising edges passing 0.5 in the processed signal trace is used to calculate the time delay between all subsequent pulses $t_n$ and hence the deviation of each individual rotation-period from the average period of this measurement. For practical reasons the terms 'jitter' and 'drift' are defined as follows (see Fig. 2). Drift is the moving average over 10 subsequent roundtrip-time deviations. RMS drift is the RMS-deviation of this drift from the mean revolution time. In many optical applications, drifts can be actively compensated for. RMS jitter is defined as the RMS-deviation of the measured $\Delta t$ from its drift and can typically not be compensated for. The bandwidth of the measured deviations is limited due to the limited number of measured revolutions, while the fastest detectable disturbance occurs at the rotation frequency itself. Therefore, these measurements do not yield information on the long-term stability (>seconds) but give valuable information on short-term effects. Measurements were carried out in air (up to 4 kHz, limited by rotor-heating) and vacuum (up to 5 kHz) for 0.1, 0.2, 0.5, 1, 2, 3, 4 and 5 kHz. For each rotation frequency 3 to 5 traces were acquired per medium, containing between 16 and 320 revolutions with varying sample-rates chosen for each rotation frequency. In addition, traces for the passively decelerating rotor just after the motor is switched off for different initial revolution speeds are measured. In Fig. 4 the transition between activated and deactivated motor drive in vacuum is shown. As soon as the motor is switched off the formerly dominant modulations disappear and an unmodulated rise of the rotation period $t$ is visible.

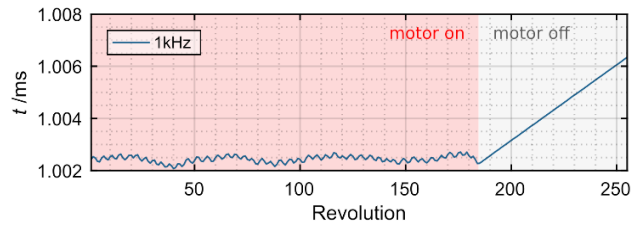

Fig. 4 Roundtrip-period deviation with activated and deactivated motor, starting at 1 kHz.

After subtracting this slope, the resulting graph shows the behavior of the free-running rotor without a driving motor (right parts of each subplot in Fig. 5). Thus, the contribution of the motor drive to fluctuations of the rotation period can be distinguished from contributions of the bearing, the surrounding medium and measurement errors. To stay as close to the original rotation-frequency as possible, all presented measurements of the free-running rotor start directly after the motor drive is switched off. Figure 6 shows the relative RMS drift and jitter values for all measurements at each rotation frequency. In vacuum, up to a rotation frequency of ~1 kHz, the relative timing jitter as well as the drift decreases with increasing frequency from ~3000 ppm to ~100 ppm. Between 1 kHz and 4/5 kHz it remains more or less constant. This holds for both the air and the vacuum measurements and with both activated and deactivated motor drive. With active motor, both the air and vacuum jitter and drift measurements are of similar magnitude. In vacuum, switching off the motor drive reduces the velocity error by ~2 orders of magnitude. The free-running measurement gives an upper limit for the experimental error of this method. At 5 kHz this is $\Delta t/t_{avg}$ = 0.85 ppm (RMS) for the jitter, and $\Delta t/t_{avg}$ = 0.15 ppm (RMS) for drifts. However, it is noteworthy that with deactivated motor jitter is dominant while for the measurements with the activated motor the drift is usually dominant. The large difference of the results with the activated and deactivated motor suggest that both jitter and drift in vacuum are mainly caused by the driving motor, while the contribution of the bearing is negligible (Fig. 5a, b; Fig. 6). With active motor drive, the fluctuation in most vacuum measurements at all rotation velocities is dominated by a modulation as illustrated in Fig. 5b. Most likely, this is due to the control loop of the rotor velocity. So far, the velocity controller is implemented in fixpoint arithmetic, and therefore the rounding error might cause this slow modulation. This effect is unexpected but might be get rid of with floating point arithmetic.

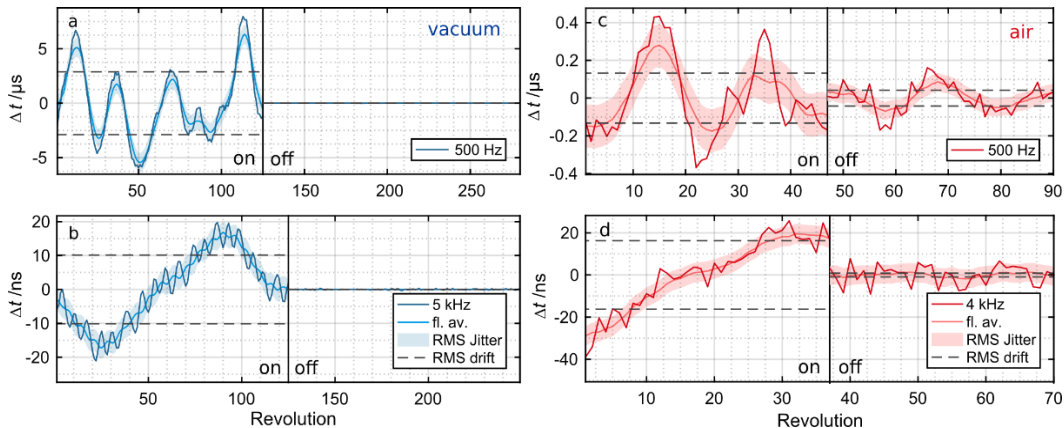

Fig. 5 Typical traces with active motor drive (left part of each subplot), and shortly after deactivation (right part of each subplot). The traces with deactivated motor are processed to remove the slope. Blue traces: Rotor in vacuum running at a) 500 Hz and b) 5 kHz, respectively. Red traces: Rotor in air running at c) 500 Hz and d) 4 kHz, respectively. Each floating average (fl. av. = drift) point is calculated taking 10 surrounding revolutions into account. Note that the y-axes of a) and c) are scaled in µs while b) and d) are scaled in ns.

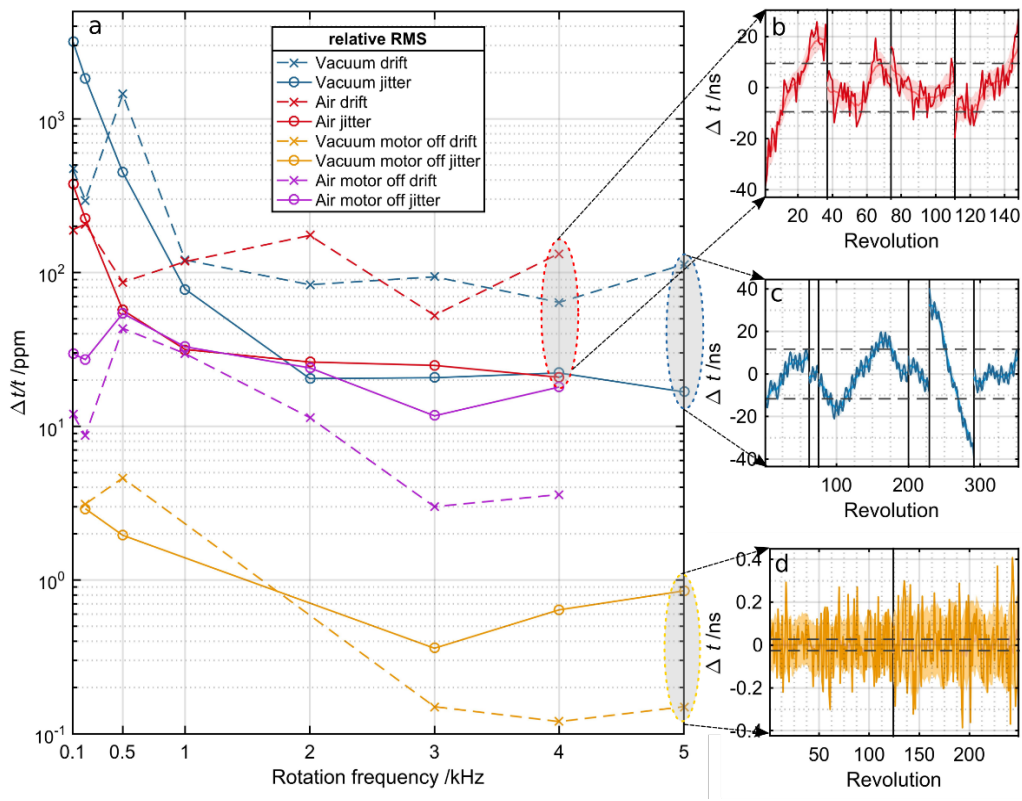

Fig. 6 a) Overview of maximum relative drift RMS and maximum relative jitter RMS for all acquired rotation frequencies in air (red), vacuum (blue), in vacuum with deactivated driver (yellow) and in air with deactivated rotor (purple). (Connecting lines between measurement points were only added to improve visibility and represent no experimental data) b) All traces of the measurements in air at 4 kHz rotation frequency (5 kHz could not be measured in air). c) All traces of the measurements in vacuum at 5 kHz rotation frequency. d) All traces of the measurements in vacuum at 5 kHz rotation frequency with deactivated motor, ring-down slope removed. The individual traces are visually separated by black lines in the plot.

In air, the magnitudes of both drift and jitter at frequencies of 2 kHz and above are similar to the vacuum measurements. The change of the rotor behavior upon deactivation of the drive, however, is distinctly different. While the drift decreases by about one order of magnitude, the jitter level is similar for most rotation frequencies. This suggests that the jitter is caused mainly by either the interaction of the rotor with the surrounding air, or air fluctuations displacing the laser beam on its path to the photodiode. In general, the traces acquired in air do not contain the fast modulations observable at all rotation frequencies in vacuum, suggesting that the contribution of the motor drive is less significant than in the vacuum measurements. At lower frequencies (100 to 500 Hz) the jitter and drift caused by the motor drive in vacuum is even higher than it is in air (see Fig. 5a/c and Fig. 6a). These differences in jitter and drift are caused by the different modulation schemes of the motor coil. During the measurement at 500 Hz in air the motor was driven with pulse-amplitude-modulation (PAM), however, at 500 Hz in vacuum the modulation was pulse-width-modulation (PWM). Although the results are not directly comparable, as both the medium and the modulation scheme is changed, this could be a hint, that PAM is more beneficial in terms of jitter and drift reduction. A further evaluation will be necessary to draw final conclusions.

## IV. Asynchronous pointing stability

The setup to measure the pointing stability is very similar to the setup described above. The photodiode is simply replaced by a high-speed camera, allowing measurements of the spatial beam displacement parallel to the rotation axis (see Fig. 7).

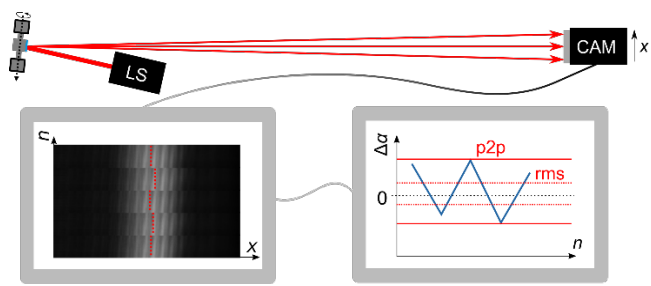

Fig. 7 LS – *Laser*: Mephisto S Non-planar Ring Oscillator (continuous wave, λ = 1064 nm). CAM – *High-Speed-Camera*: Phantom v611. Fringes are due to interference of the coherent laser light on the glass window in front of the camera sensor.

The camera allows to acquire up to $10^6$ frames per second (fps). To make sure that only one roundtrip is captured on a single frame, the frame-rate is chosen such that only every 10th frame contains an image of the beam for each rotation frequency. For every exposed frame the vertical pixels are summed up and the center of mass of these sums in the horizontal direction is calculated and recorded as the deviation from the average beam position $\Delta x$. The angular deviation was derived from the beam position on the camera via $\Delta\alpha = \arctan(\Delta x/s)$. The beam position for >9000 revolutions is monitored at rotation frequencies of 0.1, 0.2 0.5, 1, 2, 3, 4 and 5 kHz. Since the beam position is recorded at one specific rotational angle of the rotor, only the asynchronous jitter is measured, while no information about pointing jitter that is synchronous to the rotation frequency is acquired. The results of the measurement in vacuum are plotted in Fig. 8.

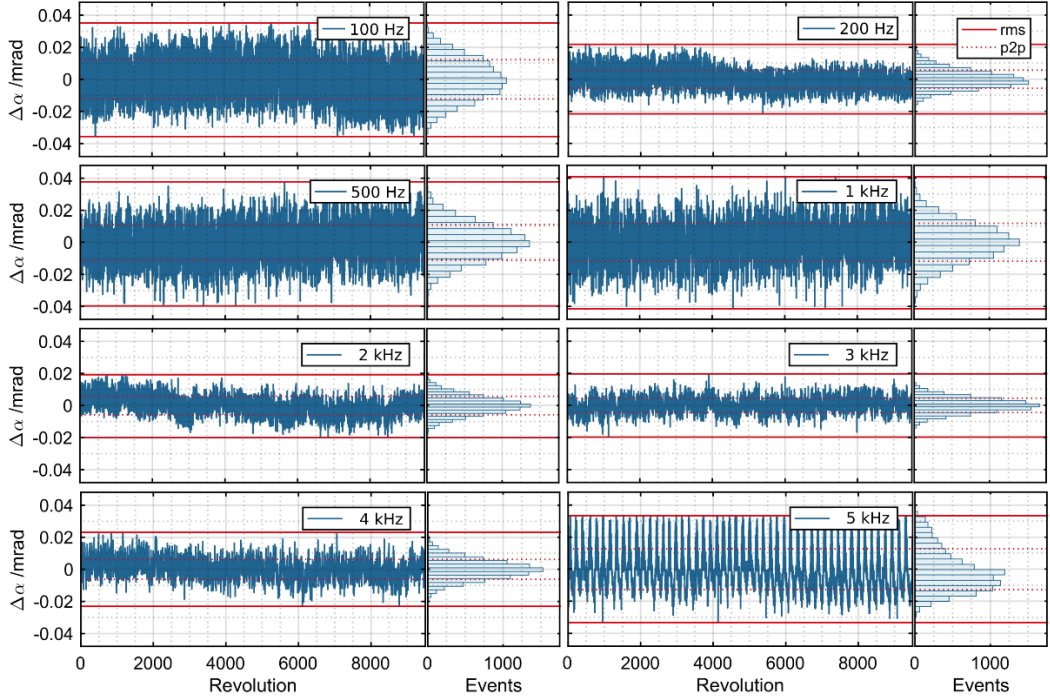

Fig. 8 Angular deviation of rotor position for >9000 subsequent revolutions at different roundtrip frequencies. The red line shows the minimum and maximum deviation and the dotted red line shows the root-mean-square deviation. Each plot also depicts the statistical distribution of each angular deviation in a histogram. The center of mass of all revolutions at one fixed round-trip frequency was taken as the position were $\Delta\alpha = 0$.

In all plots, the angular deviation from the average angle of the respective trace is shown, with all traces plotted to the same scale. The RMS angular deviations vary by a factor of ~2 for different rotation frequencies. The deviations are drastically reduced from 1 to 2 kHz. In the bearing control the position measurements are filtered with a notch filter with a corner frequency equal to the rotational speed. This notch filter is enabled for speeds higher than 1 kHz and therefore explains the reduction in the angular deviations. The traces exhibit a Gaussian distribution for all rotation frequencies except at the highest investigated frequency of 5 kHz. Here, the distribution is asymmetric with respect to its center of mass, and the trace shows a periodic temporal pattern. This pattern may be explained by the gyroscopic couplings of the rotor which are proportional to the rotation velocity. The higher the gyroscopic couplings, the lower the stability margin of the control. Figure 9 shows the offsets of the average angles for each rotation frequency together with error bars illustrating the RMS and peak-to-peak deviations.

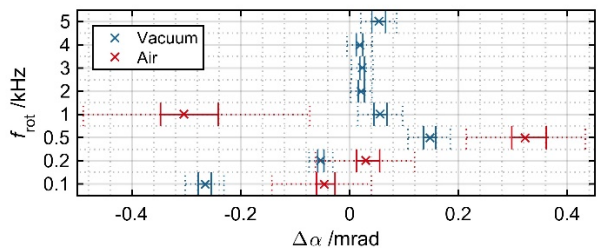

Fig. 9 Constant offset of the pointing angle and the RMS- and p2p-deviation depending on the rotation frequency for the measurements in vacuum and air.

The offset is calculated as the deviation of the center of mass for the pointing angle of each frequency from the average center of mass of all measurements. Due to changes of the setup, this is done separately for the measurements in air and vacuum and hence the absolute offset is only comparable within one medium. The offset, as well as the RMS- and p2p-deviation, are notably smaller for the vacuum measurements. This is, similar to the temporal jitter, most likely caused by air fluctuations which disturb the rotor. In air, up to 1 kHz, the change of angular offset between different rotation frequencies is far larger than the deviations within the individual traces for both the measurements in vacuum and in air (see Fig. 9). At rotation frequencies of 1 to 4 kHz, the offset of the vacuum measurement settles, and slightly changes again at 5 kHz. The offset is caused either by a static or a rotation-synchronous displacement of the rotor and is well below 400 µrad.

## V. CONCLUSION & OUTLOOK

The velocity error in form of rotation-period jitter and drift of the investigated high-speed self-bearing motor CM-AMB-400 in air and vacuum decreases significantly with increasing rotation frequency. The motor drive and its control are identified as the main sources of rotation-period jitter in vacuum with a maximum jitter of 3.4 ns (RMS) and a drift of 23 ns (RMS) at the highest revolution speed of 5 kHz. The jitter is of similar magnitude in both air and vacuum. While the rotation-period stability in air seems to be limited by air fluctuations, the measurements with deactivated motor suggest a large potential for improvement. The asynchronous beam pointing stability is significantly increased by operation in vacuum, with RMS and peak-to-peak values of 12 µrad and 33 µrad, respectively, measured over ~10000 roundtrips at 5 kHz in vacuum. The rotation-frequency-dependent angular shift may be problematic for applications which need to enable switching between different velocities without being able to readjust for the resulting offset change. The measurements presented in this manuscript allow, for the first time, to evaluate the velocity- and pointing-errors of the fastest available mechanical rotors. For the application in stack-and-dump cavities or regenerative amplifiers as discussed in [12], these results are very promising and could lead to first implementations of rotor-based switches in the near future.

ACKNOWLEDGMENTS

This work has been partly supported by the European Research Council under the ERC Grant Agreement No. [617173] "ACOPS" and by the German Federal Ministry of Education and Research (BMBF) under contract 13N13167 "MEDUSA". We thank the company Celeroton Ltd. for supplying the self-bearing permanent-magnet motor and customizing it to our needs.